\documentclass[twocolumn]{aastex6}

\begin{document}


\title{A Radio Astronomy Search for Cold Dark Matter Axions}


\author{Katharine Kelley\altaffilmark{1} and P.J.Quinn\altaffilmark{2}}
\affil{International Centre for Radio Astronomy Research (ICRAR) \\
University of Western Australia \\
Ken and Julie Michael Building \\
7 Fairway, Crawley \\
Western Australia 6009}


\altaffiltext{1}{katharine.kelley@icrar.org}
\altaffiltext{2}{peter.quinn@icrar.org}

\begin{abstract}

The search for axions has gained ground in recent years, with laboratory searches for cold dark matter (CDM) axions, relativistic solar axions and ultra-light axions the subject of extensive literature.  In particular, the interest in axions as a CDM candidate has been motivated by its potential to account for all of the inferred value of  $\Omega_{DM} \sim 0.26$ in the standard $\Lambda CDM$ model.  Indeed, the value of $\Omega_{DM} \sim 0.26$ could be provided by a light axion. We investigate the possibility of complementing existing axion search experiments with radio telescope observations in an attempt to detect axion conversion in astrophysical magnetic fields.  Searching for a CDM axion signal from a large-scale astrophysical environment provides new challenges, with the magnetic field structure playing a crucial role in both the rate of interaction and the properties of the observed photon.  However, with a predicted frequency in the radio band (200MHz - 200GHz) and a distinguishable spectral profile, next generation radio telescopes may offer new opportunities for detection.   The SKA-mid telescope has a planned frequency range of 0.4 - 13.8GHz with optimal sensitivity in the range $\sim$ 2 - 7 GHz. Considering observations at $\sim 500$MHz, the limiting sensitivity is expected to be $\sim 0.04$mK based on a 24 hour integration time.  This compares with a predicted CDM axion all-sky signal temperature of $\sim 0.04$mK using SKA Phase 1 telescopes and up to $\sim 1.17$mK using a collecting area of (1km)$^2$ as planned for Phase 2.  

\end{abstract}





\section{Introduction} \label{Intro}

In the search for dark matter, the most commonly accepted candidates are those in the group termed Weakly Interacting Massive Particles.  However with no detections to-date, attention has turned back to axions and axion-like-particles.  A detection of the cold dark matter (CDM) axion in particular would provide an elegant extension to our understanding of both cosmology and particle physics. We therefore take the opportunity with next generation telescopes coming online, to investigate the possibility of detecting observational signatures of axion CDM in and around astrophysical objects.  

The axion arose out of efforts to resolve ongoing problems in the otherwise highly successful standard model for particle physics.  One would expect a neutron electric dipole moment (edm) as a result of Charge Conjugation (C) - Parity (P) violation in the quantum field theory (QFT) for strong interactions. However, no neutron edm is observed.  Even if a small neutron edm is later found, we must ask ourselves why the associated symmetry is broken on such small scales.  \citealt{1977PhRvL..38.1440P} sought to answer this question with the introduction of a new symmetry that is spontaneously broken at very high energy. The resulting vacuum realignment mechanism in the early Universe would give rise to a real pseudoscalar particle, dubbed the axion (\citealt{1978PhRvL..40..223W, 1978PhRvL..40..279W}), whose discovery would represent a significant milestone in the development of QFT.

Initially predicted to have a mass $\gtrsim 1$keV, it was anticipated that accelerator experiments would quickly confirm the existence of this particle.  When no detections were made, a new model was required that both resolves the issue of CP violation and explains why the resulting particle has a mass $\lesssim 1$eV.  Between 1979 and 1981, models now referred to as the KSVZ (\citealt{1979PhRvL..43..103K, 1980NuPhB.166..493S}) and DFSZ (\citealt{1981PhLB..104..199D}) models were published that predict a very low mass axion with properties that are inversely proportional to the energy at which the Peccei-Quinn (P-Q) symmetry was broken, $f_{pq}$.  This damping of the mass and coupling strength by the very large P-Q energy scale is key to the axion's suitability as a candidate for CDM.  The very low mass also corresponds to a frequency in the radio band (200MHz - 200GHz), introducing the prospect of detection with radio receivers through axion-two photon coupling.

\begin{figure}
\hspace*{-2.5cm}\includegraphics[scale=0.45]{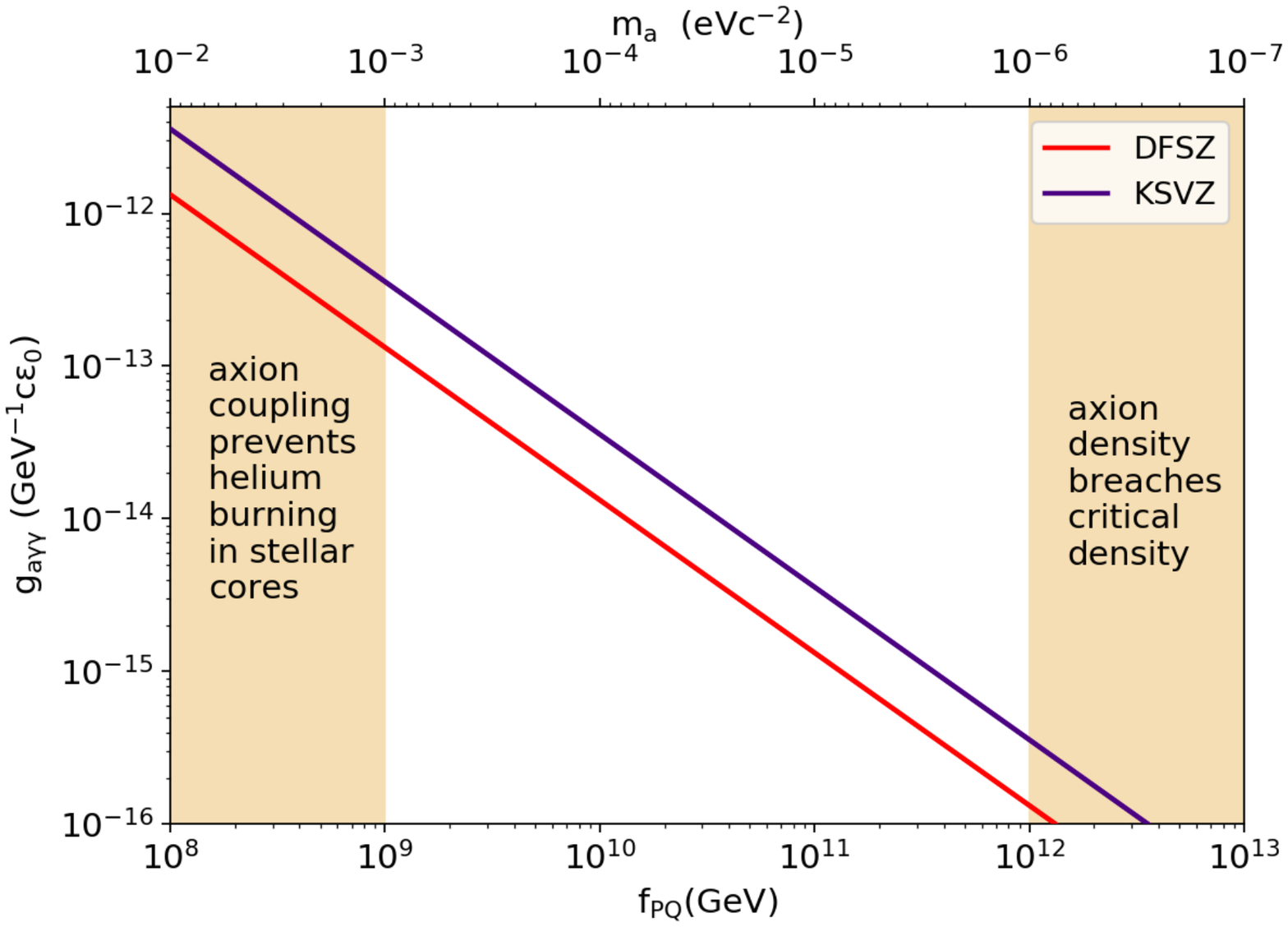}
\centering
\caption{The mass and coupling strength of the axion are determined by the P-Q symmetry breaking scale, $f_{pq}$, and constrained by astrophysical observations. The reference to KSVZ and DFSZ relates to the two generally accepted axion models from which the properties are derived.\label{coupling}}
\end{figure}

Astrophysical observations have constrained the properties of these axions. It was observed that $f_{pq}\lesssim 10^{9}$GeV would result in couplings strong enough to suppress helium burning in stellar cores.  This effect would reduce the lifetime of stars in a way inconsistent with observations (\citealt{1978PhRvD..18.1829D, 1980PhRvD..22..839D, 1987PhRvD..36.2211R}).  \citealt{1989PhRvD..39.1020B} then reinforced this constraint by observing that the duration of the neutrino burst associated with SN1987a is also consistent with a symmetry breaking scale $\gtrsim 10^{9}$GeV.  This provides an upper limit on the mass of the axion $\sim 10^{-3}$eVc$^{-2}$.  The lower limit is constrained by the critical density for dark matter (\citealt{1983PhLB..120..127P, 1983PhLB..120..137D}), see Figure \ref{coupling}.  There are a number of production mechanisms for the axion, but assuming that the minimum abundance is set by the vacuum realignment mechanism in the early Universe, we can constrain the axion mass to be above $\sim 10^{-6}$eVc$^{-2}$.

For some time axions within this parameter space were considered undetectable, their coupling to matter being extremely weak.  However, \citealt{1983PhRvL..51.1415S} proposed experimental tests to detect this `invisible' axion.  He set out two principal methods of detection; the Haloscope and the Helioscope, that have been the basis for experiments in the 35 years since. The Axion Dark Matter eXperiments (ADMX) carried out at the University of Washington and the Yale Wright Laboratory use the Haloscope approach to investigate the parameter space for axion CDM.  Early results published by \citealt{2011PhRvD..84l1302H} exclude axions in the range $3.3-3.53 \mu$eVc$^{-2}$.  And more recently \citealt{2017PhRvL.118f1302B} ruled out axions in the range $23.55 - 24.0\mu$eVc$^{-2}$.  While the axion-two photon coupling in the laboratory is the same as that expected around astrophysical objects, the transition rate in the laboratory is enhanced through the use of a resonant cavity, providing a significant advantage. However, the ability to observe extremely large volumes with a radio telescope, together with the ability to observe large parts of parameter space, makes their use highly complementary.
 
Our goal in this letter is to set out a method for computing the photon production rate for axions in the astrophysical environment and to highlight the distinguishing features for use in identifying the spectral profile.  To this end, Section \ref{Detection} of this letter sets out the photon production rate in an astrophysical magnetic field; Section \ref{Features} the distinguishing features of the resulting spectral profile, and; Section \ref{Conclusions} our conclusions.


\section{Conversion of the Cold Dark Matter Axion in an Astrophysical Magnetic Field}\label{Detection}

QFT gives rise to the prediction that an axion will couple to virtual photons provided by an electric or magnetic field, producing a single real photon (the inverse Primakoff Effect), and it is this phenomenon on which the majority of axion experiments are based.  Experiments have historically focused on two key scenarios, relativistic axions in the astrophysical environment and CDM axions in the laboratory.  Searching for CDM axions in the astrophysical environment poses new challenges, and as such we return to the original work by \citealt{1983PhRvL..51.1415S}.  He recognised that not all virtual photons from the magnetic field will contribute to axion conversion, and that the enhancement of the field in the resonant cavity together with the spatial profile of the field are crucial to detection efforts.  

In QFT, all interactions are described in terms of the emission and absorption of gauge bosons, which in the case of an electromagnetic field are virtual photons. Virtual photons do not satisfy the energy-momentum relationship $E=pc$.  We think of the moving charged particles in a particular region of space creating the magnetic field through the emission and re-absorption of virtual photons.  The energies and momenta of the virtual photons are correlated to the characteristics of the magnetic field.  Specifically, the classical magnetic field, ${\bf B}({\bf x},t)$, can be expressed as

\begin{equation}\label{Bfield}
\int{d^3{\bf k_{\gamma'}}}\int {d\omega_{\gamma'}} \ {\bf B}({\bf k}_{\gamma'},\omega_{\gamma'}) e^{i({\bf k}_{\gamma'}\cdot{\bf x}-\omega_{\gamma'} t)} 
\end{equation}

\noindent where $B({\bf k}_{\gamma'},\omega_{\gamma'})$ provides information about the number of virtual photons with energy $\hbar\omega_{\gamma'}$ and momentum $\hbar{\bf k}_{\gamma'}$. In this way it is seen that the momenta of the virtual photons are related to the spatial dependence of the magnetic field and their energies to its time dependence. Only those Fourier modes that conserve both total energy and total momentum will contribute to the conversion of the axion into a real photon, $\gamma$, by the absorption or emission of a virtual photon, $\gamma^{\prime}$.  With $\hbar{\bf k}_a\approx 0$ for axion CDM, we find from energy and momentum conservation that the magnetic field strength for the purposes of conversion is restricted to those modes with $\Delta E_{\gamma^{\prime}} \sim m_ac^2$ where $\Delta E_{\gamma^{\prime}}$ is the difference between the energy of the virtual photon and the energy, $\hbar c {\bf k}_{\gamma^{\prime}}$, it would have if it was real.  That is, energy conservation requires that $m_ac^2=\hbar k_{\gamma'}c-\hbar \omega_{\gamma'}$.

While astrophysical magnetic fields are not well understood, they have been observed in all types of galaxies and galaxy clusters, including within the Milky Way. This offers an opportunity to use next generation radio telescopes to attempt detection of axion CDM, a confirmation that would provide an elegant solution to pressing issues within both particle physics and cosmology. 

\subsection{Static Astrophysical Magnetic Fields}\label{yang}

Beginning with the cross-section determined by Sikivie for axion conversion in the laboratory, $\sigma_{lab}$, we highlight the key difference for static astrophysical fields and determine the photon production rate appropriate to conversion in the astrophysical environment.  In a static field the energy, $\hbar \omega_{\gamma'}$, of the virtual photons is zero, and the energy of the real photons produced is equal to the energy of the axion.  The cross-section in the laboratory is given by,

\begin{eqnarray}\label{Xsection}
\sigma_{lab}=\frac{c^2}{\varepsilon_0}\ \left[ \frac{1}{16\pi^2\beta_a} \right. &&\left. \left(\frac{e^2g_{\gamma}}{3\pi^2f_{pq}}\right)^2 \ \int d^3{\bf k}_{\gamma} \sum_{\lambda}\ \delta(E_a-E_{\gamma})\right. \cdot \nonumber \\
&& \left.\left|\int{d^3{\bf x}} \  {\bf B}_i({\bf x})\cdot  \epsilon_{\lambda i}^{\dagger} e^{i({\bf k}_a-{\bf k}_{\gamma})\cdot {\bf x}} \right|^2\right]
\end{eqnarray}

\noindent where the term represented in square brackets can be reconciled to Equation (6) in \citealt{1983PhRvL..51.1415S}.  The constant $c^2/\varepsilon_0$ is required to convert the cross-section to SI units and so $\sigma_{lab}$ has units of m$^{2}$.  $\beta_a$ is the velocity of the axion given by $v_a/c$, $E_a$ the energy of the axion and $\hbar {\bf k}_a$ its momentum. Likewise, $E_{\gamma}$ represents the energy of the real photon produced and $\hbar {\bf k}_{\gamma}$ its momentum.

Equation \ref{Xsection} also shows that the cross-section is proportional to the scalar product of the magnetic field vector, ${\bf B}_i({\bf x})$, and the polarisation of the real photon produced, $\epsilon_{\lambda i}^{\dagger}$.  This indicates that the interaction rate depends not only on the strength of the field but the angle between the field vector and the polarisation of the real photon.  However, in any interaction there are a number of properties that must be conserved.  One such property is parity.  \citealt{1950PhRv...77..242Y} showed that the decay of a pseudoscalar particle into two photons must result in photons with perpendicular polarisation, thus preserving the negative party of the pseudoscalar particle.  The axion is a pseudoscalar, and its conversion in a magnetic field should therefore result in a real photon with a polarisation perpendicular to the polarisation of the virtual photon.  With polarisation of a photon defined as the direction of the ${\bf E}$ field we recognise that the real photon produced will have polarisation parallel to the direction of the ${\bf B}({\bf x})$ field and 

\begin{eqnarray}
{\bf B}({\bf x})\cdot  \epsilon_{\lambda}^{\dagger} &= &|{\bf B}({\bf x})| \nonumber \\
&=&B({\bf x})
\end{eqnarray}

\noindent where the polarisation vector is a unit vector.  Lastly, $g_{\gamma}$ in Equation \ref{Xsection} is a constant $\mathcal{O}(1)$ that relates to the choice of either the KSVZ or DFSZ axion model, and all other terms have their normal meaning.

From Equation \ref{Xsection} the photon production rate can be calculated as,

\begin{equation}
\Gamma(N_a)=\sigma N_a \beta_a c
\end{equation}

\noindent where $N_a$ is the number density for CDM axions. Then,

\begin{eqnarray}
\Gamma(N_a)= \frac{c^3}{\varepsilon_0}\ \frac{1}{16\pi^2}&&\left(\frac{e^2g_{\gamma}}{3\pi^2f_{pq}}\right)^2\ N_a \int d^3{\bf k}_{\gamma} \sum_{\lambda}\ \delta(E_a-E_{\gamma}) \cdot \nonumber \\
&&\left|\int{d^3{\bf x}} \ B({\bf x})\ e^{i({\bf k}_a-{\bf k}_{\gamma})\cdot {\bf x}} \right|^2 
\end{eqnarray}

\noindent The key difference between the laboratory and astrophysical environment is that the spatial profile of the magnetic field cannot be manipulated and varies across different sources.  As such, we are not able to reduce the effect of the magnetic field in a simple way, nor do we have the ability to enhance the production rate in the same way that the resonant cavity does.  Instead, we recognise that 

\begin{equation}
\left|\int{d^3{\bf x}} \ B({\bf x})\ e^{i({\bf k}_a-{\bf k}_{\gamma})\cdot {\bf x}} \right|
\end{equation}

\noindent represents the Fourier Transform of the magnetic field. The photon production rate is therefore proportional to the value of the power spectrum associated with conservation of momentum, i.e. $B({\bf k}_{\gamma'})$ where ${\bf k}_{\gamma'}={\bf k}_a-{\bf k}_{\gamma}$.  In the astrophysical environment we can assume that the momentum of the axion is negligible, i.e. $\hbar{\bf k}_a \approx 0$, and $\hbar \omega_{\gamma'}=0$ where the field is static.  From $\Delta E_{\gamma^{\prime}} = m_ac^2$ the properties of the virtual photon must therefore be given by,

\begin{eqnarray}
k_{\gamma'}&=&k_{\gamma} \nonumber \\
&=&\frac{m_ac}{\hbar}
\end{eqnarray}

\noindent where $k=|{\bf k}|$.  The photon production rate, $\Gamma(N_a)$, in a given volume, $V$, can then be reduced to,

\begin{eqnarray}
\Gamma(N_a)\approx \frac{c^3}{\varepsilon_0}\ \frac{2}{16\pi^2}\left(\frac{e^2 g_{\gamma}}{3\pi^2f_{pq}}\right)^2\ && N_a \sum_{{\bf k}_{\gamma}} \frac{(2\pi)^3}{V} \cdot  \nonumber \\
&& \delta(E_a-E_{\gamma}) \left| B(k_{\gamma})\right|^2 \nonumber
\end{eqnarray}

\begin{eqnarray}
= \frac{c^3}{\varepsilon_0}\ \frac{2}{16\pi^2}\left(\frac{e^2 g_{\gamma}}{3\pi^2f_{pq}}\right)^2\ N_a \frac{(2\pi)^3}{V} \frac{1}{m_ac^2} && \cdot \nonumber \\
&&\left| B\left(\frac{m_ac}{\hbar}\right) \right|^2  \nonumber
\end{eqnarray}

\begin{eqnarray}\label{rate}
= 1.3 \times 10^{20} \textrm{Tesla$^{-2}$kg$^{-1}$s$^{-1}$}\left(\frac{\rho_{DM}}{V}\right) \left| B\left(\frac{m_ac}{\hbar}\right) \right|^2 
\end{eqnarray}

\noindent for an axion abundance, $\Omega_a$, equal to the critical dark matter density of $\Omega_{DM} \sim 0.26$.  There is extensive literature on the relationship between the axion mass and abundance created in the early Universe.  As the contribution to the total abundance from cosmic strings is largely unknown, we have assumed here that axions of any mass between $10^{-6} - 10^{-3}$eVc$^{-2}$ can account for the total dark matter density $\rho_{DM}$ in a given volume.

Equation \ref{rate} can then be used to assess the photon production rate in any static astrophysical field, by taking the Fourier Transform of the field profile and determining the value of $B({\bf k}_{\gamma'})$ at the required wavelength.  The majority of literature regarding astrophysical magnetic fields focuses on static spatial profiles rather than temporal fluctuations.  However, we recognise the likelihood of finding non-static fields locally within the interstellar and intra-cluster mediums and therefore consider the more general case for non-static inhomogeneous fields.

\subsection{Non-Static Astrophysical Magnetic Fields}

We have seen that the classical properties of the magnetic field are related to the energy, $\hbar \omega_{\gamma'}$, and momentum, $\hbar {\bf k}_{\gamma'}$, of the virtual photons.  A non-static inhomogeneous field is therefore expressed as,

\begin{equation}\label{Bfield2}
{\bf B}({\bf x},t)=\int {d^3{\bf k}_{\gamma'}}\int {d\omega_{\gamma'}} \ {\bf B}({\bf k}_{\gamma'},\omega_{\gamma'}) e^{i({\bf k}_{\gamma'}\cdot{\bf x}-\omega_{\gamma'} t)} 
\end{equation}
\noindent where $\hbar \omega_{\gamma'} \neq 0$. For axion CDM with $\hbar{\bf k}_a\approx 0$, we find that the magnetic field strength for the purposes of conversion is restricted to those modes with 

\begin{equation}\label{int_cons}
k_{\gamma'}c-\omega_{\gamma'} = \frac{m_ac^2}{\hbar}
\end{equation}

\noindent The photon production rate can then be determined from the action for the axion-two photon coupling,

\begin{eqnarray}
\mathcal{S}= \frac{e^2g_{\gamma}}{3\pi^2\hbar}\int d^4{\bf x}\ \frac{a}{f_{pq}}\ {\bf E}\cdot {\bf B}
\end{eqnarray}

\noindent and is written as,

\begin{eqnarray}
\Gamma(N_a)\approx \frac{c^2}{\varepsilon_0}\ \frac{1}{2}\left(\frac{e^2g_{\gamma}}{3\pi^2f_{pq}}\right)^2 && \frac{1}{T}\ \frac{1}{V}\ N_a \sum_{{\bf k}_{\gamma}}\ \frac{k_{\gamma}}{m_a} \ldotp \ \nonumber \\
&&\left| B({\bf k}_{\gamma},\omega_{\gamma}-\omega_a) \right|^2 \nonumber
\end{eqnarray}

\begin{eqnarray}
=2.2 \times 10^{25}&& \textrm{Tesla$^{-2}$kg$^{-1}$ms$^{-2}$}\ \left(\frac{\rho_{DM}}{T}\right)\ldotp \ \nonumber \\
&&\int{d^3{\bf k}_{\gamma}}\ k_{\gamma}\left| B({\bf k}_{\gamma},\omega_{\gamma}-\omega_a) \right|^2  
\end{eqnarray}

\noindent where $T$ is the time over which the temporal fluctuation of the magnetic field repeats.

 \begin{figure}
\hspace*{-1.5cm}\includegraphics[scale=0.40]{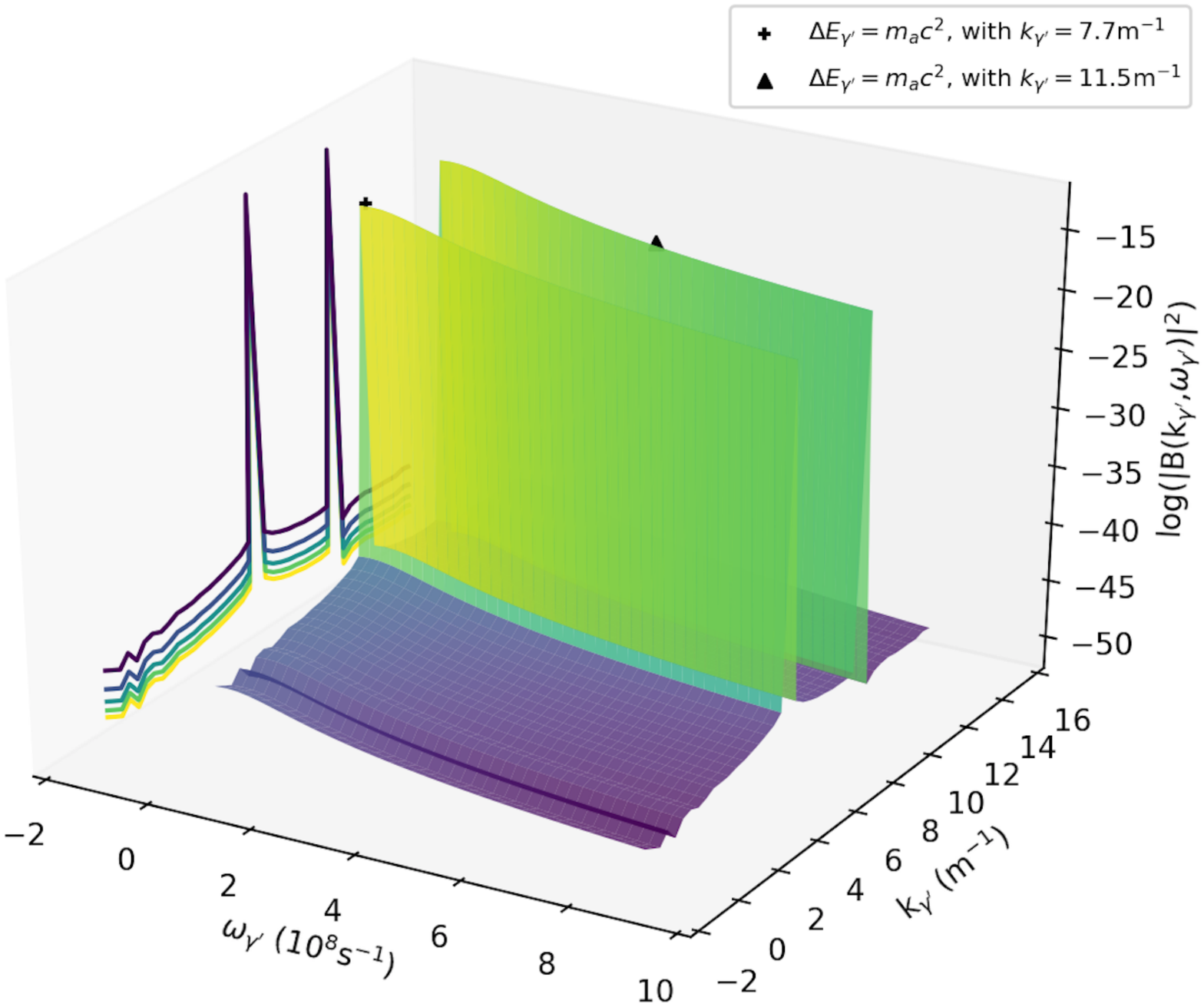}
\centering
\caption{This 2D Fourier Transform represents the spatial and temporal modes of a non-static inhomogeneous magnetic field represented by two momentum modes and a $\mu$s pulse.  An axion converting in such a field will produce real photons with two distinct frequencies, defined by the energy of the magnetic field modes contributing to the interaction.  We saw that only those modes with $\Delta E=m_ac^2$ will conserve both energy and momentum, allowing us to determine the contribution made by the field and the properties of the real photons.  The frequency of the real photons produced will be 522.6MHz and 783.9MHz based on an axion of mass $2.05\mu$eVc$^{-2}$ and momentum modes with $k_{\gamma'}=7.7$ and $11.5$m$^{-1}$. \label{pulse}}
\end{figure}

By way of example, Figure \ref{pulse} shows the 2D Fourier Transform associated with a field represented by two momentum modes and a $\mu$s pulse.  We have chosen such a distribution as a good proxy for non-static pulse fields which are available for use in the laboratory setting and for non-static fields arising from flares in the astrophysical environment.  The existence of non-static fields in the astrophysical environment will result in real photons across a range of frequencies, uniquely defined by the width of the power spectrum $\left| B({\bf k}_{\gamma},\omega_{\gamma}-\omega_a) \right|^2 $ for a particular source.  It also offers the opportunity in the laboratory to scan large areas of parameter space by fixing either the frequency or wavelength of the field. 

  
 \section{Detectability of CDM Axion Conversion}\label{Features}
 
 \begin{figure}
\hspace*{-2.5cm}\includegraphics[scale=0.45]{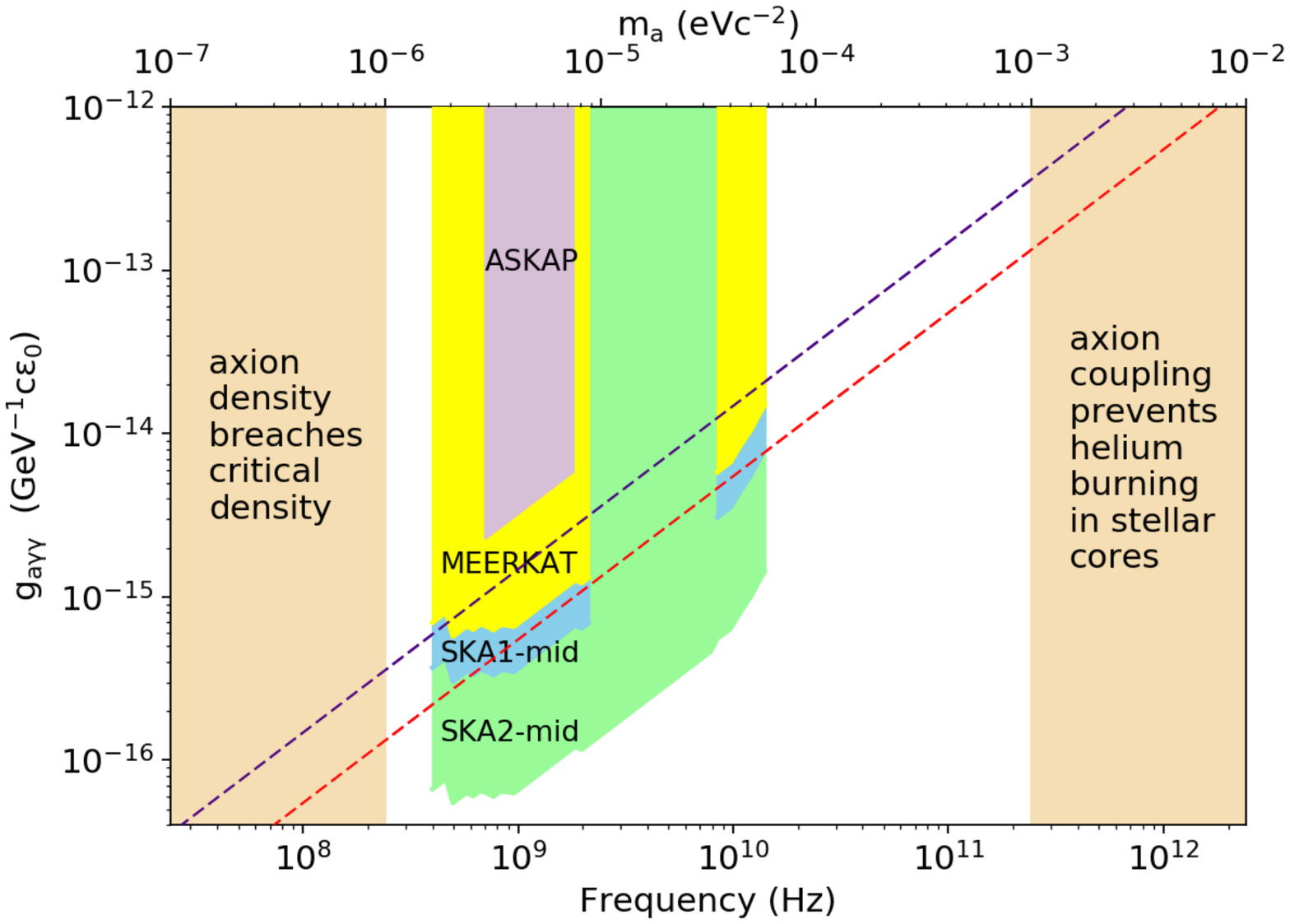}
\centering
\caption{The sensitivity of SKA-mid shows considerable improvement on the pre-cursor telescopes, the Australian SKA Pathfinder (ASKAP) and the Karoo Array Telescope (MEERKAT).  In this Figure we show the coupling strength that could be probed by observing the Interstellar Medium across the frequency range accessible to ASKAP, MEERKAT and SKA-mid.  The system temperature of the SKA is minimised between $\sim 2 - 7$GHz, corresponding to an axion mass of $\sim 8.26 - 28.91\mu$eVc$^{-2}$ and providing a good opportunity for detection of both the KSVZ and DFSZ axion. \label{sensitivity}}
\end{figure}

The spectral profile for axion conversion is influenced primarily by the mass of the axion and the relative velocities of the axion, magnetic field and observer.  Assuming an axion of mass $\sim 2.05\mu$eVc$^{-2}$ and $\hbar{\bf k}_a\approx 0$, conversion in a static magnetic field will convert to a real photon with frequency $f_{\gamma} \sim 495.6$MHz.  The width of this spectral line is then defined by the axion velocity in the frame of conversion, $h f_{gamma}\approx m_ac^2(1+v_a^2/c^2)$, and the doppler shift arising due to the relative velocities of the conversion and observer frames.  

In laboratory based experiments the magnetic field and observer reside in the same frame and the width of the spectral line is defined only by the velocity distribution of the axion.  However, in the astrophysical environment conversion occurs in a remote region where it is more difficult to assess the magnetic field's motion with respect to the observer.  We believe the most conservative position, that neither the dark halo nor the magnetic field display net rotation with respect to the rest frame of the Galaxy, is the most appropriate to take for the purposes of of our initial investigations.  In such a frame the CDM velocity distribution is given by $v_a < 300$kms$^{-1}$ and the all-sky signal will display a broadening of $\sim 200$kHz due to the orbit of the Sun around the Galactic Centre.

By assuming a simple NFW profile (\citealt{1997ApJ...490..493N}) for the CDM density, and a magnetic field strength of $50\mu$G at the Galactic Centre dissipating radially as ($r$kpc)$^{-1}$ along the Galactic disk, we can estimate the all-sky flux at a central frequency of $495$MHz to be $\sim 3.2 \mu$Jy.  In determining this flux we have assumed that the magnetic field is turbulent on small scales and that the resulting signal is spread across the surface of a sphere of radius $d$, the distance from the point of conversion to the observer.  

The higher CDM density and magnetic field strengths at the Galactic Centre make this an obvious choice of observations within the Milky Way, and with the density and magnetic field strength both dissipating radially one would expect the flux at the central frequency of $495.6$MHz to dominate the all-sky signal.  There are also additional characteristics of the Galactic Centre that may enhance this flux further. Maxwell's equations require that $\nabla \cdot {\bf B}=0$ which, when applied to Equation \ref{Bfield}, constrains the momentum vector of the virtual photon, $\hbar {\bf k}_{\gamma'}$, to be perpendicular to the direction of the classical magnetic field vector.  It is then trivial to see that in taking $\hbar {\bf k}_a \approx 0$, the direction of propagation of the real photon is perpendicular to the direction of the magnetic field vector.  With the Galactic Centre displaying coherent fields in azimuthal and z directions, this could further enhance the flux as compared to that observed along the spiral arms.

When observing such coherent fields that are perpendicular to the radial vector with Earth, in addition to the flux being maximised, the polarisation of the real photon should trace the direction of ${\bf B}$ as we saw in Section \ref{yang}.  Critically, this polarisation is perpendicular to the synchrotron radiation that will comprise a significant portion of the foreground signal.  One would also expect that observations across the sky should allow the observer to trace the properties of the magnetic field, both in strength and polarisation.  This dependence on the spatial structure of the field will be a key identifier when processing and analysing telescope data.  
 
Using the 133x15m diameter SKA1 dishes and 64x13.5m diameter MEERKAT dishes planned for SKA Phase 1, we calculate the axion antenna temperature for SKA1-mid to be $\sim 0.04$mK. The total collection area for Phase 1 is $\sim(180$m)$^2$ using this configuration, and with an assumed increase in this collecting area to ($1$km)$^2$ in Phase 2, this axion antenna temperature could increase by 2 orders of magnitude to $\sim 1.17$mK.  The ability to make such observations will be highly dependent on the sensitivity that can be achieved by SKA-mid.  Using a system temperature of 12K, a bandwidth of just 1MHz, and an integration time of 24 hours, we calculate the sensitivity limit to be $\sim 0.04$mK. Figure \ref{sensitivity} shows the full axion parameter space and the potential for observation in the Interstellar Medium using SKA-mid, the frequency range corresponding to an axion mass of $\sim 1.65$ to $57.01\mu$eVc$^{-2}$.


\section{Conclusions}\label{Conclusions}

It is clear that Earth based detection of CDM axions using radio telescopes faces significant challenges, not least in understanding the characteristics of astrophysical magnetic fields. But while there remains areas of great uncertainty, Figure \ref{sensitivity} shows that both the SKA and its precursors may offer a good opportunity for detection across large parts of parameter space.  Using simple assumptions about the properties of the CDM halo and Galactic magnetic fields, we estimate that axion conversion may result in an all-sky spectral line with  distinguishable characteristics and flux of order $1\mu$Jy.  

With significant plans underway to build next generation telescopes such as the SKA, and precursor technology becoming fully operational within 12 months, we conclude that there is a basis for both modelling and observing carefully chosen sources, an area which we will focus on in our future work.


\acknowledgments

We would like to thank Prof. Ian McArthur at the University of Western Australia for many useful discussions and invaluable guidance in preparing this work.  We also thank the referee for his valuable and constructive comments.

\end{document}